\documentclass[preprint2,numberedappendix]{emulateapj-rtx4}
\usepackage{graphicx,natbib,bm,url,color,mathtools,threeparttable,longtable}

\graphicspath{{./fig/}{./png/}}
\newcommand{\EQ}{\begin{equation}}
\newcommand{\EN}{\end{equation}}
\newcommand{\EQA}{\begin{eqnarray}}
\newcommand{\ENA}{\end{eqnarray}}

\newcommand{\EEq}[1]{Equation~(\ref{#1})}
\newcommand{\Eq}[1]{Equation~(\ref{#1})}

\newcommand{\Sec}[1]{Sect.~\ref{#1}}

\newcommand{\Fig}[1]{Figure~\ref{#1}}

\newcommand{\Tab}[1]{Table~\ref{#1}}

\newcommand{\bra}[1]{\langle #1\rangle}

{}
{}
{}

{}
{}
{}
{}
{}
{}
{}
{}
{}
{}
{}
{}
{}
{}
{}
{}
{}

{}

{}

{}
{}
{}

\newcommand{\pphi}{\hat{\bm{\phi}}}
\newcommand{\ppom}{\bm{\hat{\varpi}}}
\newcommand{\ttheta}{\bm{\hat{\theta}}}
\newcommand{\nnn}{\hat{\bm{n}}}

\newcommand{\zzz}{\hat{\mbox{\boldmath $z$}} {}}

\newcommand{\rr}{\bm{r}}

\newcommand{\bb}{\bm{b}}

\newcommand{\pom}{\mbox{\boldmath $\varpi$} {}}

\newcommand{\ii}{{i}}

\newcommand{\dd}{{\rm d} {}}

\def\degr{\hbox{$^\circ$}}

\newcommand{\G}{\,{\rm G}}

\newcommand{\MHz}{\,{\rm MHz}}
\newcommand{\GHz}{\,{\rm GHz}}

\newcommand{\s}{\,{\rm s}}

\newcommand{\uG}{\,\mu{\rm G}}

\newcommand{\cm}{\,{\rm cm}}

\newcommand{\km}{\,{\rm km}}

\newcommand{\kms}{\,{\rm km\,s}^{-1}}

\newcommand{\kpc}{\,{\rm kpc}}

\newcommand{\Gyr}{\,{\rm Gyr}}

\newcommand{\yjmp}[3]{ #1, {JMP,} {#2}, #3}
\newcommand{\yjcap}[3]{ #1, {JCAP,} {#2}, #3}

\newcommand{\yapj}[3]{ #1, {ApJ,} {#2}, #3}

\newcommand{\yapjs}[3]{ #1, {ApJS,} {#2}, #3}

\newcommand{\yana}[3]{ #1, {A\&A,} {#2}, #3}
\newcommand{\dana}[3]{ #1, {A\&A}, DOI:#2, arXiv:#3}
\newcommand{\yanas}[3]{ #1, {A\&AS,} {#2}, #3}

\newcommand{\yjetp}[3]{ #1, {Sov.\ Phys.\ JETP,} {#2}, #3}

\newcommand{\yaraa}[3]{ #1, {ARA\&A,} {#2}, #3}

\newcommand{\yprl}[3]{ #1, {PhRvL,} {#2}, #3}

\newcommand{\ymn}[3]{ #1, {MNRAS,} {#2}, #3}

\newcommand{\yprd}[3]{ #1, {PhRvD,} {#2}, #3}

\newcommand{\yjcp}[3]{ #1, {JCoPh,} {#2}, #3}
\newcommand{\yjour}[4]{ #1, {#2}, {#3}, #4}

\newcommand{\ybook}[3]{ #1, {#2} (#3)}
\newcommand{\yproc}[5]{ #1, in {#3}, ed.\ #4 (#5), #2}

\newcommand{\sana}[2]{ #1, {A\&A}, submitted, arXiv:#2}

\newcommand{\smn}[2]{ #1, {MNRAS}, submitted, arXiv:#2}
\newcommand{\sjour}[3]{ #1, {#2}, submitted, arXiv:#3}

\begin{document}

\title{Hemispheric handedness in the Galactic synchrotron polarization foreground}
\author{
Axel Brandenburg$^{1,2,3,4}$\thanks{E-mail:brandenb@nordita.org}
\& Marcus Br\"uggen$^{5}$
}

\affil{
$^1$Nordita, KTH Royal Institute of Technology and Stockholm University, Roslagstullsbacken 23, SE-10691 Stockholm, Sweden\\
$^2$Department of Astronomy, AlbaNova University Center, Stockholm University, SE-10691 Stockholm, Sweden\\
$^3$JILA and Laboratory for Atmospheric and Space Physics, University of Colorado, Boulder, CO 80303, USA\\
$^4$McWilliams Center for Cosmology \& Department of Physics, Carnegie Mellon University, Pittsburgh, PA 15213, USA\\
$^5$Hamburger Sternwarte, Universit\"at Hamburg, Gojenbergsweg 112, D-21029, Hamburg, Germany
}

\date{\!$ \, $Revision: 1.145 $ $\!}

\begin{abstract}
The large-scale magnetic field of the Milky Way is thought to be created
by an $\alpha\Omega$ dynamo, which implies that it should have
opposite handedness North and South of the Galactic midplane.
Here we attempt to detect a variation in handedness using polarization
data from the {\em Wilkinson Microwave Anisotropy Probe}.
Previous analyzes of the parity-even and parity-odd parts of
linear polarization of the global dust and synchrotron emission have
focused on quadratic correlations in spectral space of, and between,
these two components.
Here, by contrast, we analyze the parity-odd polarization itself and
show that it has, on average, opposite signs in Northern and Southern
Galactic hemispheres.
Comparison with a Galactic mean-field dynamo model shows broad qualitative
agreement and reveals that the sign of
the observed hemispheric dependence of the azimuthally averaged
parity-odd polarization is not determined by the sign of $\alpha$,
but by the sense of differential rotation.
\end{abstract}


\section{Introduction}

The main purpose of the {\em Wilkinson Microwave Anisotropy Probe}
({\em WMAP}) and {\em Planck} satellites was to map the cosmic
background radiation.
However, most of the polarized emission comes
from the Galactic foreground \citep{Adam16,Akrami18}.
Removing this contribution remains an important goal in observational
cosmology for the detection of primordial gravitational waves and
magnetic fields. 
This requires a thorough understanding of the detailed foreground
emission. The Galactic magnetic field is also of great interest to astroparticle physics,
as it is a key factor in tracing high-energy cosmic rays to their origin.
It could also be critical for understanding the hemispheric dependence
of the handedness in the arrival directions of cosmic rays \citep{KV06}
and, in particular, the gamma rays observed
with the {\em Fermi} Large Area Telescope \citep{TCFV14}.
The {\em WMAP} satellite data also allow us to learn new
important aspects about the Galaxy \citep[e.g.][]{JF12} that have never
been possible to assess systematically with conventional techniques.
In particular, Galactic synchrotron and dust polarizations can reveal
important information about the nature of its magnetic field that
can be best understood by comparing with synthetic polarization
maps form numerical simulations \citep{VGJK18}.

The determination of the magnetic field of the Galaxy is a difficult task. Most progress
has been made by using the rotation measure (RM) of pulsars or extragalactic radio
sources \citep{Haverkorn15}.
However, the large-scale pattern of the Galactic magnetic field is still largely
unknown \citep[e.g.][]{Men+08}.
\cite{Sun+08} have shown an axisymmetric disk distribution with reversals
inside the solar circle using all-sky maps at $1.4\GHz$ from the Dominion
Radio Astrophysical Observatory and the Villa Elisa radio telescope, 
the K-band map from the {\em WMAP} mission,
as well as the Effelsberg RM survey. Other efforts include the work by
\cite{Brown+07}, who used RM of extragalactic radio sources to infer an
axisymmetric pattern of the disk magnetic field.
A recent review of the models for the Milky Way 
magnetic field can be found in \cite{Boulanger+18}.

Synchrotron emission from the Galaxy dominates at low microwave
frequencies ($<30\GHz$), while thermal dust emission starts to
dominate at higher frequencies ($>70\GHz$).
Full-sky continuum maps at lower frequencies are available, for example,
at $408\MHz$ \citep{Haslam+82},
and at $1.4\GHz$ \citep{Reich+Reich86}.  \cite{Ruiz-Granados+10}
have carried out a systematic comparison of a number of Galactic
magnetic field models, which were fitted to
the large-scale polarization map at $22\GHz$.

It is believed that the Galactic magnetic field is generated by a
turbulent dynamo process, which can produce both small-scale and
large-scale magnetic fields at the same time.
Several techniques have been devised to determine signatures of
dynamo-generated magnetic fields.
One such aspect concerns the twistedness of the magnetic field
at large and small length scales.
Twist is generally quantified by magnetic and current helicities, and
various approaches have been explored to determine these quantities
\citep{VS10,JE11,OJRE11,BS14,HF14}, which are all based on Faraday
rotation.
A significant uncertainty is imposed by the fact that the polarization
data are only sensitive to the magnetic field orientation in the
plane of the sky, but not to its direction.
Under certain conditions of inhomogeneity, however, the sense of
twist can be inferred from just the polarization pattern projected on
the sky \citep{KMLK14,Bracco19,BBKMRPS19}.

Magnetic fields that are generated by an $\alpha\Omega$ dynamo
\citep{KR80} have, on average, opposite handedness
North and South of the disc plane.
It may therefore be possible to detect signatures of such a field by
analyzing the polarization patterns of the Galaxy.
Exploring this for our Galaxy is the main purpose of the present work.

The basic idea is to use the decomposition of linear polarization into
its parity-even and parity-odd parts.
In the analysis of the cosmic background radiation, one usually computes
spectral correlations between the parity-odd polarization and the
temperature.
However, as already pointed out by \cite{Bra19}, even just the parity-odd
polarization itself can sometimes be used as a meaningful proxy.
This quantity is a pseudoscalar, similar to kinetic and magnetic
helicities.
This means that it changes sign when viewing the system in a mirror.
A difficulty in applying it as a proxy for magnetic helicity is that
the parity-odd polarization is only defined with respect to a plane,
and that we can only expect a non-vanishing average if the plane is
always seen only from the same side, i.e., if one side is physically
distinguished from the other.

\section{$E$ and $B$ polarizations}

The Stokes $Q$ and $U$ linear polarization parameters change under
rotation of the coordinate system.
However, it is possible to transform $Q$ and $U$ into a proper scalar
and a pseudoscalar, which are independent of the coordinate system.
These are the rotationally invariant parity-even $E$ and parity-odd
$B$ polarizations.
They are given as the real and imaginary parts of the spherical harmonic
expansion \citep{Dur08}
\EQ
E+\ii B\equiv R=\sum_{\ell=2}^{N_\ell}\sum_{m=-\ell}^{\ell}
\tilde{R}_{\ell m} Y_{\ell m}(\theta,\phi),
\label{EBfromQU}
\EN
with some truncation $N_\ell$ and coefficients $\tilde{R}_{\ell m}$
that are related to the spectral
representation of the complex linear polarization $P=Q+\ii U$ in terms
of spin-weighted spherical harmonic functions.
They are given by \citep{Kamion97,SZ97,ZS97}
\EQ
\tilde{R}_{\ell m}=\int_{4\pi}
(Q+\ii U)\,_2 Y_{\ell m}^\ast(\theta,\phi)\,
\sin\theta\,\dd\theta\,\dd\phi,
\label{QUfromEB}
\EN
where $_2 Y_{\ell m}^\ast(\theta,\phi)$ are the spin-2 spherical
harmonics, $\theta$ is colatitude, and $\phi$ is longitude.
We choose $N_\ell=48$ for the spherical harmonic truncation.
This results in some corresponding smoothing, making it easier to discern
large-scale patterns in the resulting $E$ and $B$ polarizations.

It should be noted that \cite{ZS97} use another sign convention; see
their Equation~(6), which corresponds to a minus sign in \Eq{EBfromQU}.
Here we use Equation~(5.10) of \cite{Dur08}; see the corresponding
discussion by \cite{Bra19} and \cite{Pra+20}.

\begin{figure*}[t!]\begin{center}
\includegraphics[width=\textwidth]{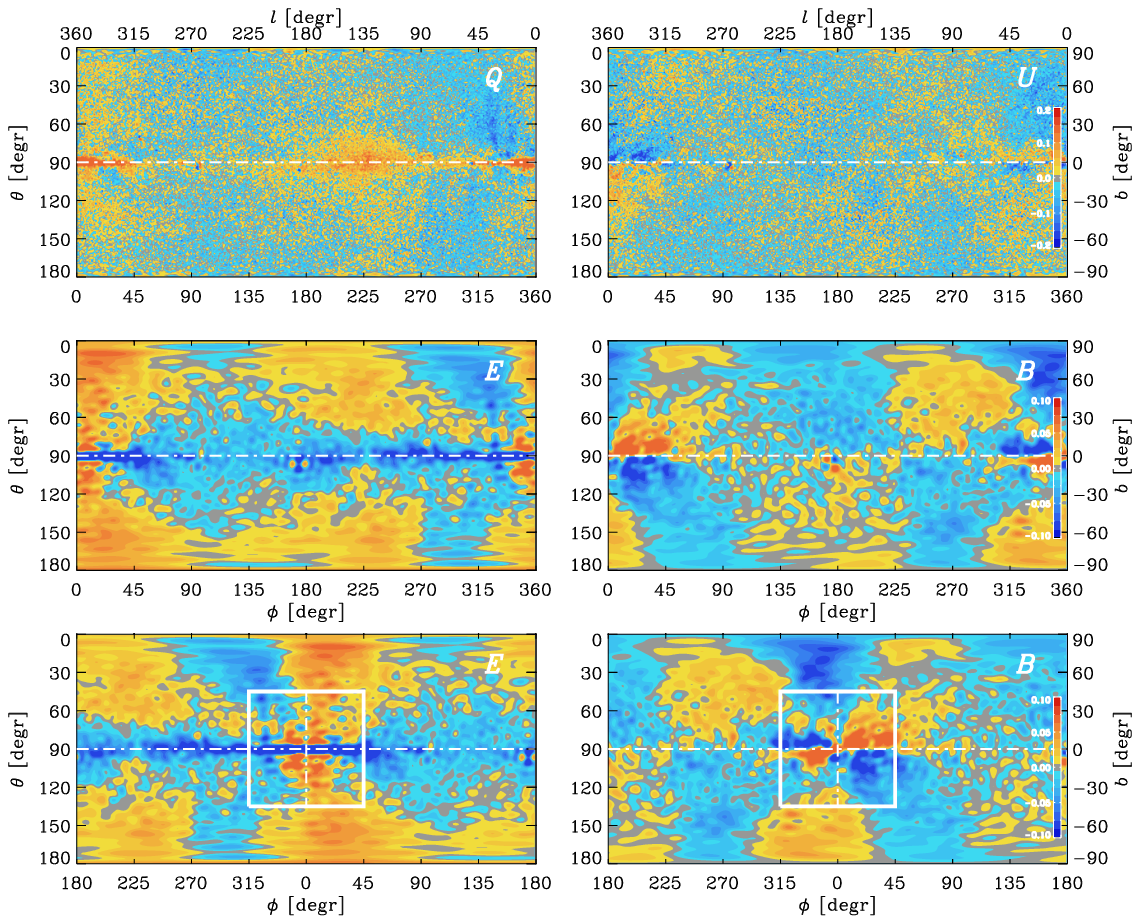}
\end{center}\caption[]{
Polarization results for the Galaxy.
{\em Top:} $Q(\theta,\phi)$ and $U(\theta,\phi)$ in mK.
Galactic $(l,b)$ coordinates are indicated on the outer axes.
{\em Middle:} $E(\theta,\phi)$ and $B(\theta,\phi)$ in mK.
{\em Bottom:} same as middle panels, but shifted such that the
Galactic center is in the middle.
Note that $\theta$ increases downward, so North (South)
is at the top (bottom).
}\label{RfromP}\end{figure*}

Following \cite{Vasil+16,Vasil+18}, we compute the spin-weighted spherical
harmonics using Jacobi polynomials $P_\ell^{(a,b)}(\cos\theta)$ as
\begin{equation}
_s Y_{\ell m}(\theta,\phi)=N_{\ell\,m}^s
\sin^a(\textstyle{\theta\over2})
\cos^b(\textstyle{\theta\over2})\,
P_{\ell-\ell_0}^{(a,b)}(\cos\theta)\,e^{\ii m\phi},
\end{equation}
where $a=|m+s|$, $b=|m-s|$, and
\begin{equation}
N_{\ell\,m}^s=(-1)^{\max(m,-s)}\sqrt{\frac
{(2\ell+1)(\ell+\ell_0)!(\ell-\ell_0)!}
{4\pi(\ell+\ell_1)!(\ell-\ell_1)!}}
\label{Norm}
\end{equation}
is a normalization factor with $\ell_0=\max(|m|,|s|)$
and $\ell_1=\min(|m|,|s|)$.
\EEq{Norm} differs from that of \cite{Vasil+18} by a factor
$\sqrt{2\pi}$ to conform with the normalization of \cite{Goldberg67}.

\section{Data selection and analysis}

Our analysis  is based on the K-band (equivalent to $22\GHz$) polarization
data obtained by the {\em WMAP} satellite after the full nine years of operation
\citep{2013ApJS..208...20B}.
This data can be downloaded from the LAMBDA website\footnote{\url
http://lambda.gsfc.nasa.gov/product/map/current} in the HEALPIX
format\footnote{\url http://healpix.jpl.nasa.gov/} \citep{Gorski+05}.

In the K-band, the emission is entirely dominated by Galactic synchrotron
emission, with spinning dust, thermal dust, and the CMB being sub-dominant
\citep{2013ApJS..208...20B}.
Thus, this band is best to study the Galactic magnetic field.

The first two panels of \Fig{RfromP} show the all-sky Stokes $Q$ and
$U$ maps at $22\GHz$ using the HEALPIX resolution parameter $N_{\rm side}=512$
(which corresponds to a pixel size of $6.8$ arcmin). We mapped this data onto a uniform grid of 
 standard spherical coordinates,
$(\theta,\phi)$, where $\theta$ is colatitude and $\phi$ is longitude,
which increases eastward (using the interpolation function in the HEALPY PYTHON package).
We also use Galactic coordinates $(l,b)$, where $l=360\degr-\phi$
is Galactic longitude and $b=90\degr-\theta$ is Galactic latitude
\citep[e.g.][]{Page+07}.
The Galactic latitude $b$ is not to be confused with the components of
the magnetic field, which will be denoted by the bold face symbol
$\bb_\perp$ so as not to confuse them with the
parity-odd constituent $B$ of the linear polarization.

\section{Results}

\subsection{Global $E$ and $B$ polarization for the Galaxy}

In \Fig{RfromP}, we show images of $Q$ and $U$ along with the rotationally
invariant counterparts $E$ and $B$ as functions of $\theta$ and $\phi$.
We see that $Q$ is mostly positive near the midplane and has maxima at
$\phi=0$ and $225\degr$.
Near $\phi=0$, $U$ is positive (negative) in the Southern (Northern)
hemisphere.
The $E$ polarization has negative extrema at $\phi\approx90\degr$
and $270\degr$, and is positive at intermediate longitudes and also at
high Galactic latitudes.
Around the Galactic center, the $B$ polarization has a characteristic
cloverleaf-shaped pattern, which is best seen in the recentered lower panels
of \Fig{RfromP}; see the white box.
This is similar to what was reported in the appendix of \cite{BF20}.
This pattern is the result of the $B$-decomposition of a purely vertical
magnetic field near the Galactic center.

\begin{figure*}[t!]\begin{center}
\includegraphics[width=\textwidth]{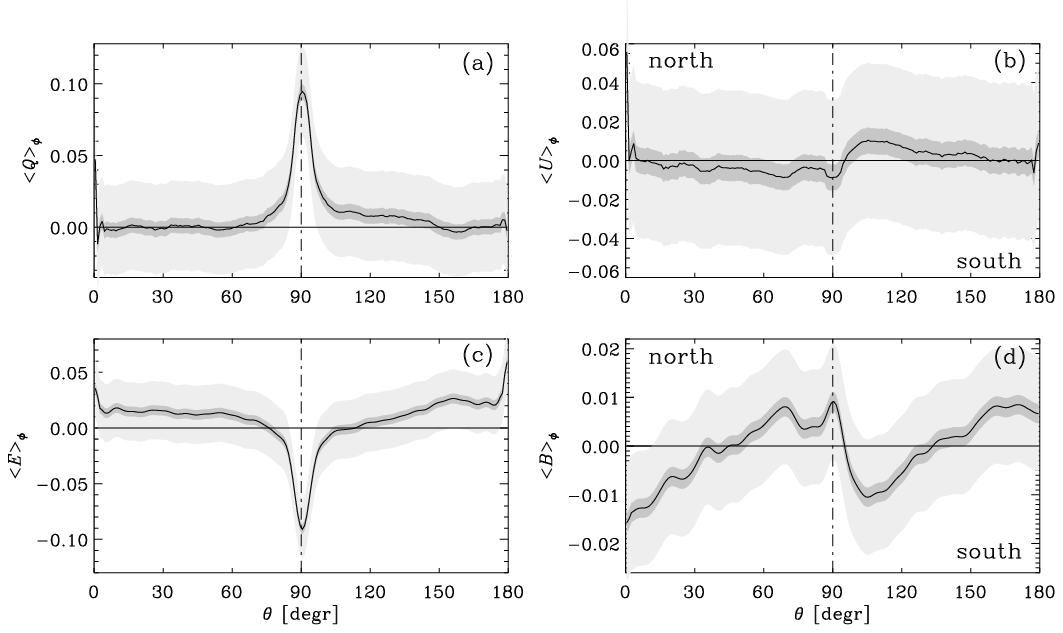}
\end{center}\caption[]{
Azimuthally averaged polarization results for the Galaxy.
{\em Top:} $\bra{Q}_\phi(\theta)$ and $\bra{U}_\phi(\theta)$.
{\em Bottom:} $\bra{E}_\phi(\theta)$ and $\bra{U}_\phi(\theta)$.
The statistical error of the mean and the standard deviation
are indicated by dark and light shades, respectively.
}\label{Rm_from_Pm}\end{figure*}

To study the systematic latitudinal dependence more clearly, we show in
\Fig{Rm_from_Pm} the $\phi$-averaged profiles of $Q$, $U$, $E$, and
$B$, which we denote by angle brackets with subscript $\phi$, i.e.,
$\bra{Q}_\phi=\int_0^{2\pi} Q\,\dd\phi/2\pi$, and likewise for
the other quantities.
We also show the standard deviation and statistical error of
the mean, where we took data that are separated by more than $10\degr$
as statistically independent.
We clearly see that, near the equatorial plane ($\theta=90\degr$),
$\bra{Q}_\phi$ and $\bra{E}_\phi$ have a similar symmetric $\theta$
dependence about $\theta=90\degr$, but with opposite signs.\footnote{
Note a different sign convention in \cite{SZ97}.}
Also $\bra{U}_\phi$ and $\bra{B}_\phi$ have opposite signs relative to
each other, but both are roughly antisymmetric about $\theta=90\degr$.
There is, however, a negative (positive) spike in $\bra{U}_\phi$
($\bra{B}_\phi$) at $\theta=90\degr$.
It may be associated with an imperfect cancelation of the
cloverleaf-shaped feature at the Galactic center.

It should be noted here that the use of azimuthal averages breaks
the rotational invariance under coordinate transformations.
This is why the azimuthal average $\bra{E}_\phi$ depends solely on
$\bra{Q}_\phi$, and $\bra{B}_\phi$ depends solely on $\bra{U}_\phi$.

\begin{table}[t!]\caption{
First few expansion coefficients of Equation~(1).
}\vspace{12pt}\centerline{\begin{tabular}{ccccccccccccccccc}
$m$ & $\mu$ & $\kappa$ &
$\kappa\tilde{E}_{2\,m}^{(\mu)}$ & $\kappa\tilde{E}_{4\,m}^{(\mu)}$ &
$\kappa\tilde{B}_{3\,m}^{(\mu)}$ & $\kappa\tilde{B}_{5\,m}^{(\mu)}$ \\
\hline
0 & K & 1000 & $56$      & $-34      $ & $\mathbf{-18}    $ & $  6     $ \\
  & Q & 5000 & $45$      & $-27      $ & $\mathbf{-35}    $ & $ 10     $ \\
  & A &  0.1 & $73$      & $-22      $ & $\mathbf{-0.038}   $ & $ -0.002 $ \\
  & B &  0.1 & $92$      & $-34      $ & $\mathbf{+0.014}   $ & $ -0.13  $ \\
  & C & 5000 & $68$      & $-27      $ & $\mathbf{-7    }   $ & $ +8     $ \\
  & D &  0.1 & $73$      & $-22      $ & $\mathbf{+0.038}   $ & $ +0.002 $ \\
1 & K & 1000 & $ 1-9\ii$ & $  4+17\ii$ & $   4+ \ii$ & $  8-5\ii$ \\
  & Q & 5000 & $ 1-8\ii$ & $    - \ii$ & $ -10-2\ii$ & $    2\ii$ \\
  & A &  0.1 & $ 1     $ & $  1      $ & $    -4\ii$ & $     \ii$ \\
  & B &  0.1 &$-41- \ii$ & $ 13      $ & $    12\ii$ & $   -5\ii$ \\
  & C & 5000 & $57+5\ii$ & $-15 -8\ii$ & $ -2-36\ii$ & $  2+9\ii$ \\
  & D &  0.1 & $ 1     $ & $  1      $ & $    -4\ii$ & $     \ii$ \\
2 & K & 1000 & $ 7-8\ii$ & $  9-22\ii$ & $-20-17\ii$ & $-13+5\ii$ \\
  & Q & 5000 & $-1-7\ii$ & $   -15\ii$ & $-13      $ & $ -9     $ \\
  & A &  0.1 & $41+3\ii$ & $ -4      $ & $  2-26\ii$ & $    5\ii$ \\
  & B &  0.1 & $28+2\ii$ & $ -5      $ & $  1-10\ii$ & $   -2\ii$ \\
  & C & 5000 & $31+4\ii$ & $ 12 -4\ii$ & $  5-50\ii$ & $-7+13\ii$ \\
  & D &  0.1 & $41+4\ii$ & $ -4      $ & $  3-26\ii$ & $ -1+5\ii$ \\
\label{Tsummary}\end{tabular}}
\tablenotemark{
The $\kappa B_{3\,0}^{(\mu)}$ are in bold.
The factor $\kappa$ is adopted for compacter notation.
}\end{table}

The full sky maps of $E$ and $B$ in \Fig{RfromP} yield a prominent
$m=2$ variation with odd symmetry about the equator.
The odd $m=0$ variation cannot be seen without azimuthal averaging.
To quantify the relative importance of the $m=0$ and $2$
contributions, we list in \Tab{Tsummary} the first few coefficients
$\tilde{E}_{\ell m}=(\tilde{R}_{\ell m}+\tilde{R}_{\ell,\,-m}^\ast)/2$ and
$\tilde{B}_{\ell m}=(\tilde{R}_{\ell m}-\tilde{R}_{\ell,\,-m}^\ast)/2\ii$,
which is opposite to the sign convention of \cite{SZ97} for $E$ and $B$.
To assess the robustness of the result, we also compare with {\em WMAP}
data in the Q-band (41\GHz).
We distinguish the two bands by superscripts K and Q;
see \cite{BB20data} for the full set of coefficients.
We also compare with simulation data (superscripts A--D)
discussed in \Sec{ResultsMFM}.
The hemispheric handedness is quantified by the coefficients
$\tilde{B}_{3\,0}^{(\mu)}$, which are negative, except for
the models $\mu={\rm B}$ and D discussed in \Sec{ResultsMFM}.

\subsection{Comparison with a Galactic dynamo model}
\label{DynamoModel}

\subsubsection{Review of the model}

To see how our results compare with a Galactic mean field dynamo model, we
analyze the model of \cite{BF20}, which was recently applied to assess
the parity-even and parity-odd polarizations for an edge-on view of the
galaxy NGC~891.
In the present work, however, we use the same model to compute a view
from the position of the Sun, located in the mid-plane $8\kpc$
from the Galactic center ($\mu={\rm A}$).
We also compare with $3\kpc$ distance ($\mu={\rm B}$), and how models with
opposite signs of the $\alpha$ effect ($\mu={\rm C}$) and both $\alpha$
and $\Omega$ ($\mu={\rm D}$).

The models have parameters similar to that of \cite{BDMSST93}, which
was designed to describe the halo magnetic field of NGC~891.
The vertical wind in the model of \cite{BDMSST93} was omitted.

We adopt a Cartesian domain of size $20\times20\times5\kpc^3$ with normal field boundary conditions.
The computations are performed with the {\sc Pencil Code} \citep{PC}
using $256\times256\times64$ meshpoints.

\begin{figure*}[t!]\begin{center}
\includegraphics[width=\textwidth]{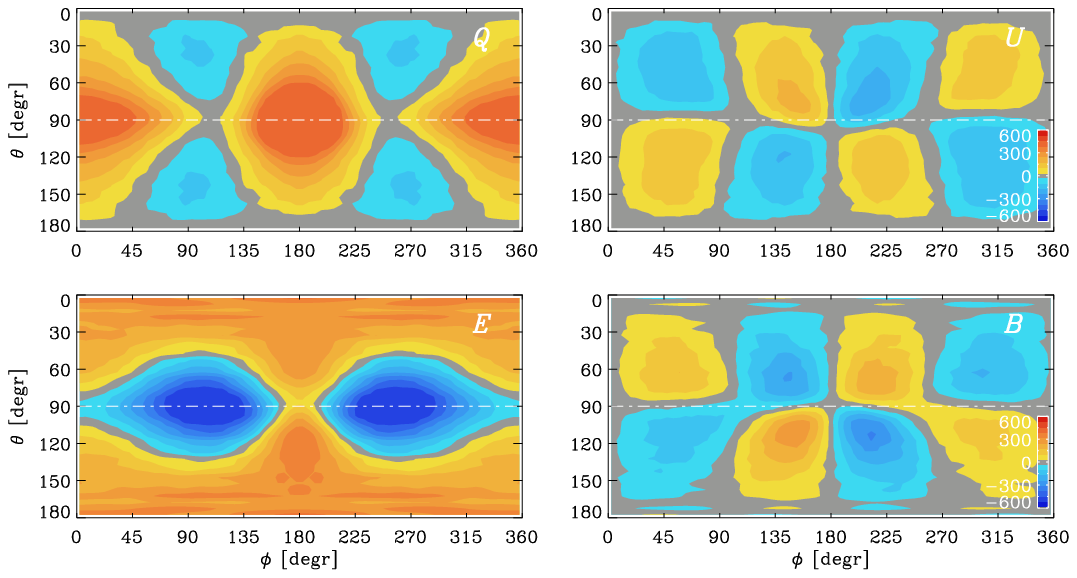}
\end{center}\caption[]{
Polarization results for the mean-field model with the Faraday
rotation term in \Eq{FarRot} included.
{\em Top:} $Q(\theta,\phi)$ and $U(\theta,\phi)$.
{\em Bottom:} $E(\theta,\phi)$ and $B(\theta,\phi)$.
}\label{EBfromQU_128x128x32_20kc_BDMSST93c_Oml}\end{figure*}

The distribution of the $\alpha$ effect in the model has a radius of
$15\kpc$ and a height of $1.5\kpc$.
In \cite{BDMSST93}, this height was associated with the thick
disk (Reynolds layer).
For the rotation curve, a Brandt profile was assumed with an angular
velocity of $\Omega_0=75\Gyr^{-1}$ and a turnover radius
of $\varpi_\Omega=3\kpc$, where the rotation law attains constant linear
velocity with $V_0=\Omega_0\varpi_\Omega\approx225\kms$.
The $\alpha$ effect has a strength of $\alpha_0=22\km\s^{-1}$
near the axis, but declines with increasing distance from
the axis and has $8\kms$ at $\varpi=8\kpc$.
It is also reduced locally by $\alpha$ quenching, which
limits the mean field to about $10\uG$.
The resulting magnetic field has quadrupolar symmetry with respect to
the midplane, i.e., the sign of the toroidal field is the same above
and below the midplane.

\subsubsection{Computation of the polarization}

To compute the apparent magnetic field from the position of the Sun in
our model, we set up a local spherical coordinate system to inspect the
local emission from a sphere around the observer at the position
$\rr_0\equiv(x_0,y_0,z_0)$ in the direction $\nnn$,
where $\nnn$ is the unit vector of $\rr-\rr_0$, with $\rr$ being the
position of a point on the sphere around the observer,
and $s=|\rr-\rr_0|$ the distance.
The cylindrical radius around the observer is $\varpi=|\pom|$, where
$\pom=(x-x_0,y-y_0,0)$, which allows us to compute the local azimuthal
unit vector as $\pphi=\zzz\times\ppom$, where $\ppom=\pom/\varpi$
and $\zzz=(0,0,1)$ are the cylindrical and vertical unit vectors,
respectively.
The third coordinate direction in our local coordinate system is
colatitude with the unit vector $\ttheta=\pphi\times\nnn$.
The polarization on the unit sphere of the observer is then
computed from $\bb_\perp=(b_\theta,b_\phi)$, whose components are
given by $b_\theta=\bb\cdot\ttheta$ and $b_\phi=\bb\cdot\pphi$.

For given wavelength, the synchrotron emissivity
is $\propto n_{\rm CR} |\bb_\perp|^\sigma$, where $\sigma\approx1.9$
\citep{GS65}.
In the following, we assume $\sigma=2$, so that the emissivity in
the complex polarization can simply be written as
\begin{equation}
P_{\rm intr}(s,\theta,\phi)=-\epsilon_0\,n_{\rm CR} (b_\theta+\ii b_\phi)^2,
\label{Pintr}
\end{equation}
where $\epsilon_0$ is a positive constant, $n_{\rm CR}$ is the cosmic
ray electron density, and $s$ is the distance from the observer.
The minus sign in \Eq{Pintr} reflects the fact that the polarization
plane represents the electric field vector of the radiation which is
orthogonal to the magnetic field in the plane of the sky.

We compute the observable complex polarization along the line of sight as
\begin{equation}
P(\theta,\phi)=\int_0^\infty P_{\rm intr}\,e^{2\ii b_r/b_{\rm F}} \,\dd s,
\label{FarRot}
\end{equation}
where $b_{\rm F}=(k_{\rm F}n_{\rm e}\lambda^2 s)^{-1}$,
with $k_{\rm F}=2.6\times10^{-17}\G^{-1}$ being a constant
\citep[e.g.][]{Pac70}, $n_{\rm e}$ the thermal electron density,
and $\lambda$ the wavelength.
Absorption can safely be neglected for our purposes.
For the sake of illustration, we adopt
Gaussian profiles for thermal and cosmic ray electron densities,
$n_{\rm e}=n_{\rm e0}\exp(-z^2/2H_{\rm e}^2)$ and
$n_{\rm CR}=n_{\rm CR0}\exp(-z^2/2H_{\rm CR}^2)$, with
$H_{\rm e}=H_{\rm CR}=1\kpc$ and midplane values
$n_{\rm e0}$ and $n_{\rm CR0}$ for the electron densities.
For $\lambda=1.36\cm$ (corresponding to $22\GHz$),
$n_{\rm e0}=0.03\cm^{-3}$, and $s=1\kpc$, we have $b_{\rm F}=230\uG$.
This is large compared to the typical Galactic magnetic field strength
of a few $\uG$, so Faraday rotation effects are weak.
We perform the line-of-sight integration by computing $P(\theta,\phi)$
on a $(\theta,\phi)$ mesh with $36\times72$ meshpoints for distances $s$
from the observer going up to $5\kpc$ in steps of $\Delta s=0.2\kpc$.

\begin{figure*}[t!]\begin{center}
\includegraphics[width=.95\textwidth]{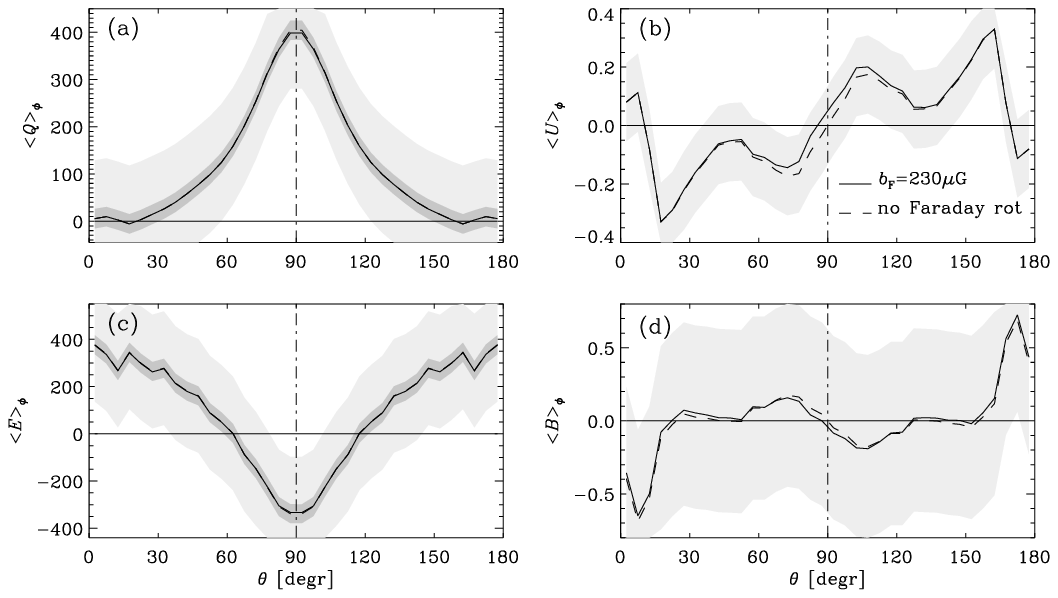}
\end{center}\caption[]{
Azimuthally averaged polarization results for the mean-field model
with Faraday rotation using $b_{\rm F}=230\uG$ (solid lines) and
without Faraday rotation (dashed lines).
{\em Top:} $\bra{Q}_\phi(\theta)$ and $\bra{U}_\phi(\theta)$.
{\em Bottom:} $\bra{E}_\phi(\theta)$ and $\bra{B}_\phi(\theta)$.
The statistical error of the mean and the standard deviation
are indicated by dark and light shades, respectively, except
for panels (b) and (d), where only the former is shown.
}\label{EBm_from_QUm_tot}\end{figure*}

\subsubsection{Results from the mean-field model}
\label{ResultsMFM}

In \Fig{EBfromQU_128x128x32_20kc_BDMSST93c_Oml} we show the results for
Model~B of \cite{BF20}.
In spite of the parameters being unrealistic for the Galaxy,
there are characteristic features that are similar to what is seen
in \Fig{RfromP} for the Galaxy: positive $Q$ at $\phi=0$ and $180\degr$,
along with negative $E$ at $\phi=90\degr$ and $270\degr$.
The results for $U$ and $B$ are not immediately obvious because there
are two nearly equally big patches of opposite sign in each hemisphere.
Only after $\phi$-averaging do we recognize a latitudinal dependence
that is similar to that of the Galaxy; see \Fig{EBm_from_QUm_tot}.
The observed sign of $\tilde{B}_{3\,0}^{(\mu)}$ emerges only when we place
the observer at a distance sufficiently far away from the Galactic center
($\mu={\rm A}$), while for smaller distances ($\mu={\rm B}$),
the sign changes and the $\bra{B}_\phi(\theta)$ profile becomes
similar to that shown in the top--right panel of Figure~1 of \cite{Bra19}.
Faraday rotation is weak, but it contributes a profile that is
symmetric about the equator.
This is because the global magnetic field has quadrupolar symmetry;
see \cite{Bra19} for a corresponding result for a dipolar field.
Changing the sign of $\alpha_0$ results in a qualitatively
different (oscillatory) dynamo, but, to our surprise, it does not
change the sign of $\tilde{B}_{3\,0}^{(\mu)}$.
The sign only changes when the differential rotation changes and thereby
the global Galactic spiral.
Changing the signs of $\alpha_0$ and $\Omega_0$ simultaneously has the
advantage of leaving the dynamo properties unchanged.

An important difference between model and observation is the fact that
in our model, the amplitude of $\bra{B}_\phi(\theta)$ is several hundred
times smaller than that of $\bra{E}_\phi(\theta)$, while for the Galaxy,
it is only about ten times smaller; see \Tab{Tsummary}.
We recall that the quadratic $EE$ correlation was found to be about
twice that of the $BB$ correlation \citep{Adam16}.
This came as a major surprise \citep{KLP17} and triggered numerous
investigations trying to explain this \citep{Kritsuk18}.
In the present paper, however, we have not studied quadratic correlations,
but the signed functions $E$ and $B$ themselves.

\section{Conclusions}

The most important result from our work is \Fig{Rm_from_Pm}(d).
Except for the spike at the equator, we see a clear hemispheric dependence
of the longitudinally averaged parity-odd $B$ polarization.
This shows that the magnetic field in North and South, which is
responsible for the polarization pattern, must be mirror
images of each other, statistically speaking.
To our knowledge, this is the first time that such a clear measurement
of handedness has ever been made for the Galaxy.
Remarkably, the results obtained for the mean-field dynamo agree
qualitatively with those for the Galaxy, although the signal is much
weaker in the model; see \Fig{EBm_from_QUm_tot}(d).

To interpret our results further, we must learn how to decipher the
$B$ signal.
There is no one-to-one correspondence between $B$ polarization and
magnetic helicity.
Indeed, $B$ can be zero even for fully helical turbulence
\citep{BBKMRPS19,Bracco19}.
We only obtain a finite signal if one viewing direction is preferred
over another, as was discussed in those two papers.
Whether or not this argument actually works for the Galaxy is not a priori clear
because the $\alpha$ effect produces magnetic helicity of opposite
signs at large and small scales.
In the Sun, for example, observations of active regions tend to reveal
only the small-scale contribution \citep{Pra+20}.
Our present paper now shows that this may be different for the Galaxy.
\cite{BCEB12} emphasized, however, that much of the Galactic polarized
emission is caused by a turbulent anisotropic component, which \cite{JF12}
called stiated.
It is therefore plausible that the detected hemispheric handedness
is caused by the opposite orientations of the Galactic spiral in the
two hemispheres; see the reversed sign of $\tilde{B}_{3\,0}^{(\mu)}$
for $\mu={\rm D}$ in \Tab{Tsummary}.

Our paper reveals a number of other properties in
the $(\theta,\phi)$ maps of $E$ and $B$ that also agree qualitatively
with the dynamo model: negative extrema of $E$ at the equator near
$\phi=90\degr$ and $270\degr$, and two positive (negative) extrema of
$B$ at $\phi=45\degr$ and $225\degr$ in the North (South).
However, the sign of the azimuthally-averaged $B$ appears not to be
related to the sign of the $\alpha$ effect, as was originally hoped,
but it seems to reflect the spiral nature of the Galaxy.
Looking South gives a mirror image of the Galactic spiral compared to
the view towards North.
This new finding is supported by considering the quantity
$\tilde{B}_{3\,0}$ in our models.

\acknowledgements

We thank Andrea Bracco for having verified the hemispheric dependence
of $\bra{B}_\phi$ after we posted our preprint.
We also thank Rainer Beck, Anvar Shukurov and Kandaswamy Subramanian
for suggesting useful improvements to the paper.
This work was supported in part through the Swedish Research Council,
grant 2019-04234, and the National Science Foundation under the grant
AAG-1615100 (AB).
MB acknowledges support from the Deutsche Forschungsgemeinschaft
under Germany's Excellence Strategy -- EXC 2121 ``Quantum Universe''
-- 390833306.
This work was in part performed at the Aspen Center for Physics, which
is supported by National Science Foundation grant PHY-1607611.
We acknowledge the allocation of computing resources provided by the
Swedish National Allocations Committee at the Center for Parallel
Computers at the Royal Institute of Technology in Stockholm.


\end{document}